\documentclass[prb,aps,twocolumn]{revtex4}
\usepackage{graphicx,color,latexsym}
\usepackage{dcolumn}
\usepackage{amsmath,amssymb,epsf,bm}
\begin{document}
\title{Spontaneous generation of vortices by a nonuniform Zeeman field in a two-dimensional Rashba coupled superconductor}
\author{A.~G. Mal'shukov}
\affiliation{Institute of Spectroscopy, Russian Academy of Sciences, Troitsk, Moscow, 108840, Russia}
\affiliation{Moscow Institute of Physics and Technology, Institutsky per.9, Dolgoprudny, 141700 Russia}
\affiliation{National Research University Higher School of Economics, Myasnitskaya str. 20, Moscow, 101000 Russia}
\begin{abstract}
The interplay of the electron exchange interaction and spin-orbit coupling results in spontaneous supercurrents near magnetic insulator islands, which are placed on the top of a two-dimensional (2D) superconductor, and whose magnetization is parallel to 2D electron gas.  It is shown that in contrast to the well studied situation, where such an effect involves only topologically trivial spatial variations of the superconducting order parameter, one should take into account supercurrent vortices. The latter are spontaneously generated around the island's boundary of an arbitrary shape and result in screening of the Zeeman field. This problem has been considered for electrons subject to a strong Rashba spin-orbit coupling, including Dirac systems as well. In the latter case vortices can carry Majorana zero modes.
\end{abstract}
\maketitle

\section{Introduction}

The Zeeman interaction of Cooper pair electrons with  strong magnetic or exchange fields produces a depairing effect. It destroys the superconducting state \cite{Chandrasekhar,Clogston}, or makes the superconducting order parameter nonuniform \cite{Fulde,Larkin}. The spin-orbit coupling (SOC) of  electrons substantially modifies the effects of the Zeeman field. One of the most striking manifestation of the interplay between this field and SOC is the magnetoelectric effect \cite{Edelstein, Yip}, which takes place even at a weak Zeeman interaction. It results in a spontaneous supercurrent, which was predicted  \cite{Malsh island,Pershoguba,Hals} to circulate near a magnetic insulator island deposited on the top of a 2D superconductor or a thin film whose electrons are subject to the Rashba \cite{Rashba} SOC. At the same time, the Zeeman field is oriented parallel to the 2D electron gas. In spatially uniform systems this effect results in a helix spatial variations of the  order parameter \cite{Edelstein,Samokhin,Barzykin,Agterberg,Kaur,Agterberg2,Dimitrova}. In this case the electric currents, which are created by the magnetoelectric effect and by the order-parameter phase gradient, compensate each other, so that the net current is zero. In contrast, for the varying in space Zeeman field such a compensation is absent, which results in a finite supercurrent  \cite{Malsh island,Pershoguba,Hals}. Basically, the same magnetoelectric effect leads to the anomalous Josephson effect in the so called $\phi_0$ junctions \cite{Krive,Reinoso,Zazunov,ISHE,Liu,Yokoyama,Konschelle,Assouline} where the Zeeman field and SOC inside the junction induce the spontaneous Josephson current, even in the absence of the external phase difference between superconducting contacts.

In thermal equilibrium the spatial variations of the order parameter must minimize the free energy of the electronic system which interacts with a Zeeman field. So far, the theory  was mostly restricted to situations where the order parameter has  a trivial topology. At the same time, it is important to understand the role of topologically nontrivial spatial configurations of the order parameter, such as vortices. As known, vortices may be induced in type II superconductors under the orbital effect of an external magnetic field. Therefore, it is natural to ask, whether the exchange interaction of electron spins with a proximized magnetic insulator could induce vortices. For example, the creation of single edge vortices which can carry Majorana zero modes was considered \cite{Malsh Maj} in a special case of a Zeeman field produced by a long  thin magnetic wire on top of a 2D Dirac superconductor. In this connection it is important to consider  magnetic islands of various shapes and to find out general conditions for the formation of vortices. Moreover, a  theory is needed not only for topological superconductors, but also for spin-orbit coupled superconductors having the trivial topology. Since the number of vortices and their positions strongly depend on the strength and orientation of the Zeeman field, their interaction with magnetic excitations opens new possibilities to control  hybrid system involving superconducting and magnetic components, with great potential for practical applications.

It will be shown below that a chain of vortices  may be generated along the border of a magnetic  island, as well as at a domain wall.  At the same time, supercurrents, which are created by such chains, screen the currents that are produced by the direct magnetoelectric effect and in some cases completely compensate them in the bulk of an island and outside it, irrespective of its shape.

The parameters of such a vortex  state are calculated by minimizing the free energy functional which, in turn, will be obtained using the semiclassical theory of Green's functions \cite{Larkin semiclassical,Eilenberger,Rammer,Serene} for disordered superconductors. Two models  will be considered: a 2D superconductor with the Dirac electron band and a 2D superconductor with the parabolic band and the strong Rashba SOC. It is assumed that the electronic band splitting, which is associated with this SOC, is much larger than the superconducting gap, while the latter is larger, or comparable to the Zeeman field. Within the chosen model the s-wave superconductivity in the 2D gas has either an intrinsic origin, or is induced by the proximity effect. The model is supported by recent experimental works on 2D superconductors with strong spin-orbit effects. For example, some reports claim that the Dirac surface band migrates from the surface of a three-dimensional topological insulator (TI) towards the top of a thin superconducting over-layer and acquires a superconducting gap \cite{Sedlmayr,Trang}. Self-formation  of a thin superconducting Rashba-coupled film at the interface of the Pb superconductor and TI was observed  \cite{Bai}, a 2D superconducting film of Tl, Pb, Bi and In was grown on a semiconductor substrate  \cite{Yoshizawa,Matetskiy,Haviland,Zhang,Uchikashi,Sekihara}. The strong SOC coupling and  superconductivity was observed also in insulator's interface. \cite{Pavlov, Caviglia} The symmetry of the superconducting state in these systems is not known. Therefore, their relevance to the considered model has yet to be found out.
\begin{figure}[tp]
\includegraphics[width=6cm]{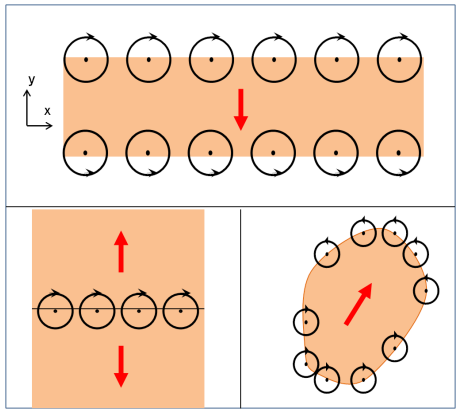}
\caption{(Color online)  A chain of supercurrent vortices is spontaneously formed at the border of a region with a nonzero Zeeman/exchange interaction. The region has a form of a long strip (top), or an arbitrary island (bottom right). Vortices also appear at a domain wall (bottom left). The magnetization direction is depicted in each case by thick arrows} \label{fig1}
\end{figure}

\section{Free energy}

The analysis  is based on the free energy functional $\mathcal{F}$ which depends on the order-parameter of the superconductor $\Delta=|\Delta|\exp i\theta$. This functional is derived from the one-particle Hamiltonian. For a 2D superconductor the latter has the form
\begin{eqnarray}\label{H}
&&H=\tau_3\frac{\hat{k}^2}{2m}+\tau_3\alpha(\sigma^x \hat{k}_y-\sigma^y \hat{k}_x)+\bm{\sigma}\mathbf{Z}-\tau_3\mu  + \nonumber\\
&&\tau_3V_{\text{imp}}+\mathrm{Re}[\Delta]\tau_1-\mathrm{Im}[\Delta]\tau_2\,,
\end{eqnarray}
where $\mu$ is the  chemical potential,  $\mathbf{Z}=(Z_x,Z_y)$ is the Zeeman field produced by the exchange interaction with spins of the magnetic island, $\mathbf{\hat{k}}=-i\partial/\partial\mathbf{r}-(e/c)\tau_3\mathbf{A}$, and $\bm{\sigma}$ is a vector composed of Pauli matrices $\sigma^j$  ($j=x,y,z$).  $\mathbf{A}$ is the vector-potential of the magnetic field which is induced by the supercurrent. The Pauli matrices $\tau_i$, $i=1,2,3$, operate in the Nambu space. The second term  in Eq.(\ref{H}) represents the Rashba SOC. In the case of Dirac electrons, such as electrons on the surface of a three dimensional (3D) TI, the first parabolic term should be ignored. In the following, the sufficiently strong Rashba coupling $\alpha$ will be assumed, so that the  splitting $\Delta_{\text{so}}=2 \alpha k_F$ of the electron energy is much larger than $ |\Delta|$, where $k_F$ is the Fermi wave vector.  The term $V_{\text{imp}}$ in Eq.(\ref{H}) describes the random impurity potential. It will be assumed that the mean elastic scattering rate  $1/\tau_{\text{imp}}$ of electrons on impurities is much larger than $|\Delta|$ and $Z$. At the same time $1/\tau_{\text{imp}} \ll \Delta_{\text{so}}$. The latter condition can take place in some 2D systems formed at interfaces, where heavy atoms are involved. For example, $\Delta_{\text{so}}$ as large as 0.1eV is expected in Pb films on the GaAs substrate \cite{Sekihara}.

The free energy will be calculated within  the semiclassical approximation which is valid when $\mu \gg |\Delta|,Z, 1/\tau_{\text{imp}}$ and $|\bm{\nabla} Z| \ll k_F Z$.  The Eilenberger-Usadel formalizm  \cite{Larkin semiclassical,Eilenberger,Rammer,Serene} may be employed  in this parameter range to derive semiclassical equations for the Green function. Furthermore, due to the relatively large $\Delta _{\text{so}}$ these equations may be projected onto two spin-splitted bands \cite{Houzet}, or on a single band in the case of TI \cite{Zyuzin,Bobkova,Linder}. It is important that the  low temperature $k_B T \ll |\Delta|$ is assumed. Therefore, the free energy  is different from the Ginzburg-Landau functional, which is valid near the critical temperature where $|\Delta| \ll k_BT$. However, one can use the fact that  $|\Delta|$ strongly varies only near vortex cores, where it turns to zero, while it  stays approximately constant in the most of the system. In contrast, the supercurrent which is associated with the phase gradient $\bm{\nabla}\theta$ decreases slowly outside the core. Especially, in thin films the vortex size is large because the screening effect of the induced magnetic field is weak \cite{Pearl}.  As long as the size of the vortex is much larger than the superconductor's coherence length $\xi$, a contribution of the core in the vortex energy may be neglected. Therefore, in the following $|\Delta| $ will be fixed equal to the spatially uniform bulk value $\Delta_0 > 0$. Such an approximation implies a weak depairing effect of the Zeeman field. According to Ref. \cite{Houzet},  in a strongly Rashba coupled system this effect can be ignored at $Z \ll \sqrt{\Delta/\tau_{\text{imp}}}$.

At fixed $|\Delta|=\Delta_0$ and  $k_BT \ll  \Delta_0$ one may calculate the free energy by expanding semiclassical Green's functions over $\bm{\nabla}\theta$ and $Z$, up to quadratic terms. As  shown in Appendix A, the free energy is given by
\begin{equation}\label{F}
\mathcal{F}=\frac {\pi}{4} N_FD\Delta_0 \int d^2r  \left((\bm{\nabla}\theta-2\frac{e}{c}\mathbf{A}+\kappa\mathbf{F})^2  +\beta F^2\right)\,,
\end{equation}
where $F_x=-2Z_y/v_F$ and $F_y=2Z_x/v_F$.  $N_F$ and $D$ are the state density and diffusion constant of 2D electrons, respectively. $\kappa$ and $\beta$ are dimensionless parameters. In a 2D Rashba system $D=\tau_{\text{imp}}(\mu/m+\alpha^2)$ \cite{Houzet},  $N_F=m/2\pi$, $\beta=2(1-r^2)/(1+r^2)^2$ and $\kappa=2r/(1+r^2)$, where $r=\alpha/\sqrt{2\mu/m+\alpha^2}$. At the same time, in a Dirac system $D=2\alpha^2 \tau_{\text{imp}}$, $N_F=\mu/2\pi \alpha^2$,$\beta=0$  and  $\kappa=1$.


\section{Simple examples}

\subsection{Infinite strip}

Let us assume that $Z$ is finite inside a strip which is infinite in the $x$-direction and has the width $w$. $\mathbf{Z}$ is directed parallel to the $y$-axis, as shown in Fig.1. Therefore, $F_y=0$ and $F_x>0$. If $w \ll \lambda_s$, where $\lambda_s$ is the effective magnetic screening length, one may ignore $\mathbf{A}$ in Eq.(\ref{F}). Note that in thin films this length is much larger than the London penetration depth \cite{Pearl}. By varying Eq.(\ref{F}) with respect to $\theta$ we arrive to the equation
\begin{equation}\label{phase}
\nabla^2\theta + \kappa\bm{\nabla}\mathbf{F}=0
\end{equation}
In the considered geometry a solution of this equation is $\theta=0$. The phase is finite only near remote edges of the magnetic strip at $|x|\rightarrow \infty$ where $\bm{\nabla}\mathbf{F} \neq 0$ \cite{Malsh island}. As a result, the integrand in Eq.(\ref{F}) becomes $(\kappa^2  +\beta) F^2$. The first term originates from the kinetic energy of the supercurrent which is induced by the magnetoelectric effect, while the second term represents a direct depairing effect of the Zeeman field.

However, the above solution of Eq.(\ref{phase}) is not unique. One may consider vortex solutions, where $\theta$ winds up an integer of $2\pi$ around singular points. Usually, vortices increase the superconductor's energy. In contrast, in the considered situation they can reduce it, thus leading to a spontaneous formation of vortices around the island. Let us consider two regular chains of vortices with the period $d \gg \xi$, which are placed along the boundaries of the strip at $y=-w/2$ and $y=w/2$, respectively, where $\xi \ll w$. It is assumed that counterclockwise and clockwise vortices with the vorticity 1 are placed on the opposite boundaries. In this case $\theta$ in Eq.(\ref{F}) is given  by the sum of phases $\theta_i=\phi(\mathbf{r}-\mathbf{r}_i)$, where $\phi(\mathbf{r}-\mathbf{r}_i)$  is the angle between $\mathbf{r}-\mathbf{r}_i$ and the $x$-axis for the $i$-th vortex. Since $Z$ is small, its effect on the coordinate dependence of $\theta_i$ is neglected. It is convenient to represent $\bm{\nabla}\theta$ as
\begin{equation}\label{nablathetabar}
\bm{\nabla}\theta=\overline{\bm{\nabla}\theta}+\bm{\delta\theta} \,,
\end{equation}
where $\overline{\bm{\nabla}\theta}$ is the average of $\bm{\nabla}\theta$ over $x$, while $\bm{\delta\theta}$ is the deviation from the average. As shown in Appendix B, $\overline{\bm{\nabla}\theta}_y=0$, while  $\overline{\bm{\nabla}\theta}_x=2\pi/d$ at $-w/2<y<w/2$, and $\overline{\bm{\nabla}\theta}_x=0$ elsewhere. By denoting the first term in Eq.(\ref{F}) as $\mathcal{F}_{\text{kin}}$ and substituting there Eq.(\ref{nablathetabar})  we arrive  to
\begin{equation}\label{Fkin}
\mathcal{F}_{\text{kin}}= \frac{\pi}{4} N_FD\Delta_0 \left(Lw(\kappa F-2\pi/d)^2+\int d^2r(\delta\theta)^2\right)\,.
\end{equation}
It is seen that at $\kappa F=2\pi/d$ the first term vanishes, while at $2\pi w \gg d$ the second term, as shown in Appendix B material, plays the role of the line energy, because $\delta\theta$ is not small only within a shell $\sim d$ near the island's boundary.  As a result, in the leading order with respect to $d/w$ one may take into account only the first term in Eq.(\ref{Fkin}). Hence,  the formation of the vortex "coat" at the boundary of the magnetic island becomes energetically favorable. The vortex lattice period is given by  $d=2\pi/\kappa F= \pi v_F/\kappa Z$. With decreasing  $w$ the second term in Eq.(\ref{Fkin}) must be taken into account. Fig.2 demonstrates that in this case  the lattice period $d$ deviates from its asymptotic ($w\rightarrow \infty$) value $d= \pi v_F/\kappa Z$.
\begin{figure}[tp]
\includegraphics[width=6cm]{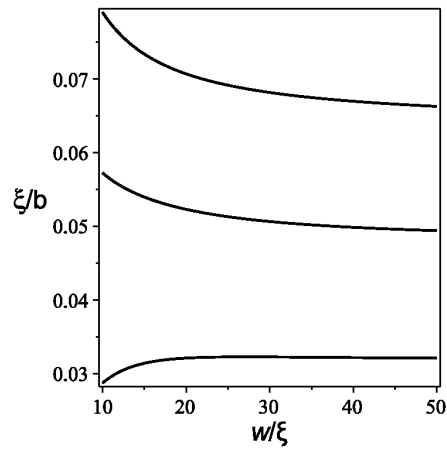}
\caption{(Color online)  Dependence of the inverse distance between vortices  on the strip width, at various strengths of the Zeeman field. From top to bottom: $\xi \kappa Z/v_F=0.2, 0.15$  and $0.1$.} \label{fig2}
\end{figure}

\subsection{Arbitrary island's shape}

The above theory may be extended to the case of an arbitrary island's shape. Let us assume that the island is much smaller than the magnetic screening length and ignore $\mathbf{A}$ in Eq.(\ref{F}). In Eq.(\ref{phase}) the phase may be written in the form $\theta=\theta_v+\theta_F$, where $\theta_F$ is associated with the magnetoelectric effect, while $\theta_v$ originates from vortices. The former  turns to 0 at $F=0$. In contrast, $\theta_v$ does not vanish with $F$ and satisfies the equation $\nabla^2\theta_v=0$.
Let us denote $\bm{\nabla}\theta+\kappa \mathbf{F}=\mathbf{J}$. From this definition one obtains
\begin{equation}\label{curlj}
\bm{\nabla}\times\mathbf{J}=\bm{\nabla}\times(\kappa\mathbf{F}+\bm{\nabla}\theta_v)\,.
\end{equation}
Note, that the second term on the right is not zero, because $\theta_v$ is singular in the vortex center. In contrast, the nonsingular $\bm{\nabla}\times(\bm{\nabla}\theta_F)=0$. Therefore, it does not enter in Eq.(\ref{curlj}). The current conservation (Eq.(\ref{phase})) gives $\bm{\nabla}\mathbf{J}=0$. Therefore, $\mathbf{J}$ may be represented  in the form $\mathbf{J}=\bm{\nabla}\times\mathbf{B}$, where $\mathbf{B}$ is parallel to the $z$-axis. By substituting the so expressed $\mathbf{J}$ into Eq.(\ref{curlj}) we arrive to
\begin{equation}\label{B}
\nabla^2B_z=-(\rho_F+\rho_v)\,,
\end{equation}
where $\rho_F=\kappa[\bm{\nabla}\times\mathbf{F}]_z$, $\rho_v=[\bm{\nabla}\times\bm{\nabla}\theta_v]_z$ and the subscript $z$ denotes the vector's $z$-component. $\rho_F$ plays the role of the "magnetic charge", which is responsible for the magnetoelectric effect. At the same time,  $\rho_v$ may be called the "vortex charge" density. Integration of $\rho_v$ over a small area enclosing a vortex gives  $2\pi l$, where the integer $l$ is the vorticity. Therefore, $\rho_v=2\pi \sum_i l_i\delta(\mathbf{r}-\mathbf{r}_i)$, where $\mathbf{r}_i$ is the position of the $i$-th vortex. If the distance between vortices is much smaller than the size of an island one can average $\rho_v$ over $\mathbf{r}_i$. It is important that one can always choose the averaged vortex spatial distribution $\overline{\rho_v}$ in such a way that $\rho_F+\overline{\rho_v}=0$ in Eq.(\ref{B}). As a result $\overline{B_z}=0$, as well as the averaged $\overline{\mathbf{J}}=0$. Hence, by ignoring fluctuations of $\rho_v$, the first term in Eq.(\ref{F}) which can be written as $\overline{J}^2$, turns to zero. Therefore, the free energy is minimized by an appropriate choice of  the vortex distribution. If the Zeeman field is uniform inside an island,  the magnetic and vortex charges are concentrated on its boundary. With decreasing island's size the line energy given by the second term in  Eq.(\ref{Fkin}) becomes important and  the employed above macroscopic approach fails. In this case a few vortices can help to reduce the free energy \cite{Malsh Maj}.

\subsection{Domain wall}

If the Zeeman field is not uniform inside an island, the magnetic charge $\rho_F=\kappa[\bm{\nabla}\times\mathbf{F}]_z=2\kappa(\bm{\nabla}\cdot\mathbf{Z})/v_F$  is finite there. As an example, let us consider a domain wall  in Fig.1, where $\mathbf{Z}$ is directed in the positive $y$-direction for $y>0$ and vice versa for $y<0$. The width of the wall is sufficiently small to ignore its internal structure. In this case  $\rho_F=\kappa[\bm{\nabla}\times\mathbf{F}]_z= -2\kappa F_x\delta(y)$. Hence, in Eq.(\ref{B}), $\rho_F+\overline{\rho_v}$ turns to zero, if a row of counterclockwise vortices with the density $\kappa F_x/\pi$ is placed along the domain wall. Similar to previously discussed examples, such a vortex distribution minimizes the free energy. If the size of the island in the $y$-direction is finite, the magnetic charges with the line density $\rho_F=\kappa F_x/2\pi$ reside on both boundaries together with two compensating chains of vortices.

\section{Magnetic screening}

Usually, the magnetic screening  is produced by supercurrents which induce the magnetic field  directed in the opposite direction to the external field. As a result, the magnetic  field can not penetrate deep into superconductors. In 3D samples it decreases exponentially together with the screening current. However, in thin films such a screening effect is weak. It leads only to a $r^{-2}$ reduction of the current at $r\gtrsim \lambda_s$ \cite{Pearl}.  A similar situation takes also place in the case when the supercurrent and vortices are induced by the considered here magnetoelectric effect of the Zeeman (exchange) field. The screening effect  must be taken into account for  magnetic islands whose size is comparable with $\lambda_s$. The magnetic field is represented in Eq.(\ref{F}) by $\mathbf{A}$. This vector-potential satisfies the Maxwell equation
\begin{equation}\label{A}
\nabla^2\mathbf{A}=-\frac{4\pi}{c}\delta(z)\mathbf{j}\,.
\end{equation}
For a disordered superconductor the 2D electric current density is given by $\mathbf{j}=e\pi|\Delta| N_F D(\bm{\nabla}\theta+\kappa \mathbf{F}-2(e/c)\mathbf{A})$. By substituting this current in Eq.(\ref{A}) the vector-potential may be expressed in terms of $\mathbf{J}=\bm{\nabla}\theta+\kappa \mathbf{F}$. As a result, instead of $\mathbf{J}-2(e/c)\mathbf{A}$, that enters in  the free energy Eq.(\ref{F}), we obtain the  expression whose Fourier transform is given by
\begin{equation}\label{Aq}
\mathbf{J}_{\mathbf{q}}-2(e/c)\mathbf{A}_{\mathbf{q}}=\mathbf{J}_{\mathbf{q}}\frac{q}{q+k_s}\,,
\end{equation}
where $k_s=4\pi^2 e^2|\Delta| N_F D/c^2$. It is seen that at the large $q$ we arrive to an unscreened expression ($\mathbf{A}=0$), while at $q\lesssim k_s$ a crossover occurs from $1/r$ to $1/r^2$ spatial dependence of the vortex current \cite{Pearl}. A typical $k_s^{-1}$ is larger than 1$\mu$m.

Let us take into account in Eq.(\ref{F}) only $\bm{\nabla}\theta$ and $\mathbf{A}$ which are averaged over vortex positions.  The total energy $\mathcal{F}_{\text{tot}}$ must include also the energy density $(rot \mathbf{A})^2/8\pi$ of the induced magnetic field. By expressing $\overline{\mathbf{A}}$ in terms of $\overline{\mathbf{J}}$ and by integrating the magnetic energy over $z$ we arrive to
\begin{equation}\label{F2}
\mathcal{F}_{\text{tot}}= \frac{\pi}{4} N_FD\Delta_0 \sum_\mathbf{q}  \frac{|\mathbf{\overline{J}}_{\mathbf{q}}|^22q^3+|\mathbf{q}\times\mathbf{\overline{J}}_{\mathbf{q}}|^2k_s}{2q(q+k_s)^2} \,.
\end{equation}
It follows from this equation that $\mathcal{F}_{\text{tot}}$ turns to zero if $ \mathbf{\overline{J}}_{\mathbf{q}}=0$. It is the same condition which regulates the formation of vortices in the absence of the screening effect.  At the same time, the line energy, which is given by the second term in Eq.(\ref{Fkin}), was ignored in Eq.(\ref{F2}). It is reasonable to expect, however, that the screening effect can not modify the surface energy, if the distance $d$ between vortices is much smaller than $k_s^{-1}$, because the surface term in Eq.(\ref{Fkin}) is determined by a thin shell $\sim d $ near the boundary. Anyway, the screening effect is not important for large enough islands where the surface energy constitutes  only a small fraction.

\section{Conclusion}

The spatial configuration of the superconducting order parameter in a 2D spin-orbit coupled superconductor with an inhomogeneous Zeeman interaction  has been considered. The Zeeman field was assumed to be directed parallel the 2D electron gas. Such a field may be produced by the exchange interaction of 2D electrons with polarized electrons of a magnetic insulator film which is placed on top of the superconductor. It is shown that a chain of vortices is spontaneously formed on the  island's  boundary, as well as along a domain wall, even at a weak Zeeman interaction. The distance between vortices increases for a weaker Zeeman field and decreases down to the coherence length when this field becomes comparable to the order parameter. This situation is quite different from the special island's geometry which was considered in Ref.\cite{Malsh Maj}, where the Zeeman field must be sufficiently strong to create a pair of vortices with localized Majorana zero modes at the ends of a long magnetic wire. It is shown that the formation of the vortex "coat" takes place for sufficiently large islands having an arbitrary shape. It is important that such a coat suppresses the supercurrent which is induced by the Zeeman interaction inside the island and outside it, except for a thin shell along the boundary, or a domain wall.

It should be noted that the considered model is restricted to 2D superconductors and very thin films where SOC and the Zeeman/exchange interaction, which presumably are  interface effects, can affect  the superconducting order parameter. This situation is different from the popular model where the superconductivity in a 2D normal system is induced by a contact to a massive 3D superconductor. In the latter case (for example in Ref.[\onlinecite{Fu}])  the order parameter, which may incorporate vortices, originates from the superconductor and does not depend on 2D physics.

A most important  application of the discussed effect of the Zeeman/exchange field is related to topological superconductors where vortices carry  Majorana zero modes (MZM). \cite{Fu,Jackiw,Read,Ivanov} By varying the direction of the Zeeman field it is possible to vary positions and the number of vortices and, hence, of MZM's which are bound to them. For example, in the case of the strip in Sec.III (Fig.1), the density of  vortices is controlled by $Z_y$. Therefore, the interaction between MZM's can be varied by rotation of $\mathbf{Z}$. Also, MZM's which are localized at magnetic domain walls can be manipulated by moving these walls. One more interesting problem arises in connection with the topologically nontrivial magnetization of  the adjacent magnetic insulator which can lead to a finite magnetic "charge" in Eq.(\ref{B}). This charge is localized at the center of a magnetic vortex or a skyrmion. As it follows from Eq.(\ref{B}) and is discussed in Sec.III B, the supercurrent vortex may be localized at the core of such a magnetic structure which, in turn,  may be manipulated by certain means.

This work has been focused on an analysis of equilibrium properties of the considered hybrid systems. On the other hand, one would expect many interesting nonequilibrium phenomena which are associated with an interaction of vortices and magnons in a magnetic insulator film. Such an interaction results, as was just discussed, in motion of vortices along the island boundary. In turn, moving vortices generate the electric field and lead to the finite resistivity of the superconductor, as it usually takes place in the vortex state of type II superconductors. Such effects may result in new functionalities of superconducting quantum circuits with integrated magnetic components.

Vortices, which are induced by the Zeeman/exchange field at the magnetic island boundary might be directly observed by means of the surface tunneling microscopy, which is usually employed in connection with conventional vortices. For example, this method was applied to In monolayers on Si in Ref.[\onlinecite{Yoshizawa}].

\emph{\textbf{Acknowledgements}} - The work was partly supported by the Russian Academy of Sciences program "Actual
 problems of low-temperature physics."


\appendix

\section{Calculation of the free energy}

The free energy in Eq.(\ref{F})  is given by the second-order expansion with respect of $\bm{\nabla}\theta,\mathbf{A},Z$ and the first order expansion in their cross-products. It is convenient to transform Hamiltonian Eq.(\ref{H}) with the help of the unitary operator $\exp( i\tau_3\theta/2)$, which results in the Hamiltonian
\begin{eqnarray}\label{Htilde}
&&\tilde{H}=\frac{\tau_3}{2m}(\hat{\mathbf{k}}+\tau_3\bm{\mathcal{A}})^2+\tau_3\alpha[\sigma^x (\hat{k}_y+\tau_3\mathcal{A}_y)-\nonumber\\
&&\sigma^y (\hat{k}_x+\tau_3\mathcal{A}_x)]+\bm{\sigma}\mathbf{Z}-\tau_3\mu  + \tau_3V_{\text{imp}}+\Delta\tau_1\,,
\end{eqnarray}
where $\bm{\mathcal{A}}=\bm{\nabla}\theta/2-(e/c)\mathbf{A}$ and $\hat{\mathbf{k}}=-i\partial/\partial\mathbf{r}$. Note, that the above unitary transformation could pose definite problems, which are associated with the singularity of $\theta$ in the presence of vortices \cite{Franz}.  These problems, however, are mostly related to an analysis of quasiparticle states, while our task is to get an expansion of the free energy over gradients of the order parameter, far from the cores of vortices. Formally, within the semiclassical approximation this expansion can be obtained with or without the gauge transformation, both approaches leading to the same result.

In Eq.(\ref{Htilde}) $\Delta$ is chosen real and positive. It depends strongly on coordinates close to the vortex cores. However, sufficiently far from them, at the distance much larger than the coherence length this dependence is weak. At small $\mathcal{A}$  and $Z$ one may neglect their effect on $\Delta$ and take the latter equal to $\Delta_0$, which is the unperturbed order parameter. $\tilde{H}$ can be represented in the form $\tilde{H}=\tilde{H}_0 +\delta H$, where $\tilde{H}_0$ is the unperturbed Hamiltonian and the interaction Hamiltonian is given by
\begin{equation}\label{deltaH}
\delta H=\frac{1}{2m}(\hat{\mathbf{k}}\bm{\mathcal{A}}+\bm{\mathcal{A}}\hat{\mathbf{k}}+\tau_3\bm{\mathcal{A}}^2)+\alpha(\sigma^x \mathcal{A}_y-\sigma^y \mathcal{A}_x)+\mathbf{Z}\bm{\sigma}
\end{equation}
Let us define $\tilde{H}_{\lambda}=\tilde{H}_0 +\lambda\delta H$. Then, the correction to the free energy can be represented as \cite{AGD}
\begin{equation}\label{FA}
\delta \mathcal{F}=-\frac{k_BT}{2}\sum_{\omega_n}\int_0^1 d\lambda\int d^2r\mathrm{Tr}[\delta HG_{\lambda,\omega_n}(\mathbf{r,r})]\,,
\end{equation}
where the trace is taken over Nambu and spin variables and $G_{\lambda,\omega_n}(\mathbf{r,r}^{\prime})$ is the imaginary-time Green function which is calculated with the Hamiltonian $\tilde{H}_{\lambda}$. This function should be calculated up to the first-order with respect to $\delta H$.

First we consider the Rashba spin-orbit coupling (SOC) in a topologically trivial 2D system.  It will be assumed that the splitting of electron energy bands $\Delta_{so}$, which is caused by the SOC,  is much larger than $\Delta_0$ and the elastic scattering rate $1/\tau$. In this case it is convenient to transform $\tilde{H}$ to the helical basis. In this basis, instead of spin projection, the quantum states are labeled by the helicity index $\nu=\pm$. Accordingly, there are two bands with the energies $\xi_{\mathbf{k}}^{\nu}=k^2/2m + \nu \alpha k - \mu$. Each band crosses the Fermi energy ($\xi_{\mathbf{k}}^{\nu}=0$) at the corresponding wave-number $k_F^{\nu}$. The Fermi velocity $v_F=\sqrt{2\mu/m + \alpha^2}$ is the same in both bands, while densities of states $N^{\nu}_F=(m/2\pi)(1-\nu\alpha/v_F)$ are different. Therefore, there are different electron scattering rates $1/\tau^{\pm}=2\pi N_F^{\pm}\overline{|V_{\text{imp}}|^2}$, where $\overline{|V_{\text{imp}}|^2}$ is expressed in terms of the Born  scattering amplitude for random uncorrelated impurities \cite{AGD}.

The calculation of $\delta \mathcal{F}$ in Eq. (\ref{FA}) can be simplified, if the  semiclassical Green function is used instead of $G_{\lambda,\omega_n}(\mathbf{r,r}^{\prime})$. The semiclassical approximation is valid when the Fermi wave length  $\sim k_F^{-1}$ is much less then other relevant parameters having the dimension of length. In the considered case of the strong SOC there are two helical bands. Therefore, semiclassical Green functions are defined for each band, while their nondiagonal terms, which correspond to mixing between bands, may be ignored at the large $\Delta_{so}=|\xi_{\mathbf{k}}^{+}-\xi_{\mathbf{k}}^{-}|$. The semiclassical functions are defined in the following way:
\begin{equation}\label{g}
g^{\nu}_{\lambda,\omega_n}(\mathbf{n}_{\mathbf{k}},\mathbf{r})=i\frac{\tau_3}{\pi}\int d\xi_{\mathbf{k}}^{\nu} G_{\lambda,\omega_n}(\mathbf{k,r})\,,
\end{equation}
where  $\mathbf{n}_{\mathbf{k}}=\mathbf{k_F}/k_F$ and $G_{\lambda,\omega_n}(\mathbf{k,r})$ is obtained  from $G_{\lambda,\omega_n}(\mathbf{r,r}^{\prime})$  by Fourier transform with respect to $\mathbf{r}-\mathbf{r}^{\prime}$, and by setting $(\mathbf{r}+\mathbf{r}^{\prime})/2 \rightarrow \mathbf{r}$. Note, that the  semiclassical function depends only on the direction of $\mathbf{k}$, which in turn is fixed on the Fermi line. Semiclassical equations for $g^{\nu}_{\lambda,\omega_n}(\mathbf{r})$ in the presence of the strong Rashba SOC and magnetic field have been derived in Ref.[\onlinecite{Houzet}]. These results will be employed for the calculation of the free energy Eq.(\ref{FA}). In terms of quasiclassical functions Eq. (\ref{FA}) can be  written in the form
\begin{eqnarray}\label{FA2}
&&\delta \mathcal{F}=-i\frac{\pi k_BT}{4}\sum_{\omega_n}\int d\mathbf{n}_{\mathbf{k}}\int d^2r\int_0^1 d\lambda\times \nonumber \\ &&\mathrm{Tr}[\tau_3(v_F\mathbf{n}_{\mathbf{k}}\bm{\mathcal{A}}+\mathbf{Z}\bm{\sigma})g_{\lambda,\omega_n}(\mathbf{n}_{\mathbf{k}},\mathbf{r})]\,,
\end{eqnarray}
where
\begin{eqnarray}\label{gA2}
&&g_{\lambda,\omega_n}(\mathbf{n}_{\mathbf{k}},\mathbf{r})=\frac{N_F^{+}}{2}g_{\lambda,\omega_n}^{+}(\mathbf{r})(1+\bm{\sigma}\times\mathbf{n}_{\mathbf{k}})+\nonumber\\
&&\frac{N_F^{-}}{2}g_{\lambda,\omega_n}^{-}(\mathbf{r})(1-\bm{\sigma}\times\mathbf{n}_{\mathbf{k}})\,.
\end{eqnarray}
It is easy to see that due to the angular averaging only the odd in $\mathbf{k}$ part of $g$ contributes in the integral Eq. (\ref{FA2}). In the case of the strong impurity scattering the Green function is almost isotropic. Therefore, it can be represented as $g^{\nu}_{\lambda,\omega_n}(\mathbf{n}_{\mathbf{k}},\mathbf{r})=g^{\nu}_{\lambda,\omega_n}(\mathbf{r})+\mathbf{n}_{\mathbf{k}}\mathbf{g}^{\nu}_{\lambda,\omega_n}(\mathbf{r})$ . \cite{Usadel} The second term, which represents the anisotropic correction, is small, so that $|\mathbf{g}| \ll 1$. From Eqs.(\ref{FA2}-\ref{gA2}) one thus obtains
\begin{eqnarray}\label{FA3}
&&\delta \mathcal{F}=-iv_F\frac{\pi k_BT}{8}\sum_{\omega_n}\int d^2r\int_0^1 d\lambda\times\nonumber\\
&&\mathrm{Tr}_{\tau}[\tau_3(N_F^{+}\mathbf{g}^{+}_{\lambda,\omega_n}+N_F^{-}\mathbf{g}^{-}_{\lambda,\omega_n})2\bm{\mathcal{A}}+\nonumber \\
&&\tau_3(N_F^{+}\mathbf{g}^{+}_{\lambda,\omega_n}-N_F^{-}\mathbf{g}^{-}_{\lambda,\omega_n})\mathbf{F}]\,,
\end{eqnarray}
where $F_x=-2Z_y/v_F$ and  $F_y=2Z_x/v_F$. As shown in Ref.[\onlinecite{Houzet}], the functions $\mathbf{g}^{\pm}_{\lambda,\omega_n}$ can be expressed in terms of the isotropic function $\tilde{g}_{\lambda,\omega_n}(\mathbf{r})=(1/2)(g^{+}_{\lambda,\omega_n}(\mathbf{r})+g^{-}_{\lambda,\omega_n}(\mathbf{r}))$. The corresponding equations have the form
\begin{eqnarray}\label{gplusA}
&&N_F^{+}\mathbf{g}^{+}_{\lambda,\omega_n}-N_F^{-}\mathbf{g}^{-}_{\lambda,\omega_n} =-v_FN_F\tau^3\tilde{g}_{\lambda,\omega_n}\times\nonumber \\ &&\left(\frac{1}{\tau_{+}^2}\bm{\nabla}_{+}\tilde{g}_{\lambda,\omega_n}-\frac{1}{\tau_{-}^2}\bm{\nabla}_{-}\tilde{g}_{\lambda,\omega_n}
+\frac{i\lambda}{\tau_{+}\tau_{-}}\mathbf{F}[\tau_3,\tilde{g}_{\lambda,\omega_n}]\right)\nonumber \,,\\
&&N_F^{+}\mathbf{g}^{+}_{\lambda,\omega_n}+N_F^{-}\mathbf{g}^{-}_{\lambda,\omega_n} =\nonumber \\ &&-v_FN_F\tau^3\tilde{g}_{\lambda,\omega_n}\left(\frac{1}{\tau_{+}^2}\bm{\nabla}_{+}+\frac{1}{\tau_{-}^2}\bm{\nabla}_{-}\right)\tilde{g}_{\lambda,\omega_n} \,,
\end{eqnarray}
where $N_F=(N_F^++N_F^-)/2$, so that $\tau^{\pm}/\tau=N_F/N_F^{\pm}$. The gradients $\bm{\nabla}_{\pm}$ are given by
\begin{equation}\label{nablaA}
\bm{\nabla}_{\pm}*=\bm{\nabla}*+\frac{\lambda i}{2}(\bm{2\mathcal{A}}\pm \mathbf{F})[\tau_3, *]\,.
\end{equation}
As can be seen from Eq.(\ref{FA3})  it is sufficient to calculate the  linear in $\mathcal{A}$ and $Z$ correction to $\tilde{g}$.  The equation for this function has been derived in Ref.[\onlinecite{Houzet}]. An analysis of this equation shows that the linear correction is absent, if one takes into account Eq.(\ref{phase}). Therefore, one may set $\tilde{g}$  equal to the zero-order Green function $\tilde{g}_{\lambda,\omega_n}=(\tau_3\omega_n+\tau_2\Delta_0)/\sqrt{\omega_n^2+\Delta_0^2}$ and ignore the gradient in the right-hand side of  Eq.(\ref{nablaA}). By substituting so simplified Eq.(\ref{nablaA}) into Eq.(\ref{gplusA}) and Eq.(\ref{FA3}) we arrive to the free energy given by Eq.(\ref{F}). The sum over $\omega_n$ was calculated at the low temperature $k_BT \ll \Delta_0$.

In the case of a Dirac system with a single Dirac point there is only a single helical band, which should be taken into account at sufficiently large $\mu$.  In this respect, the above algebra is simplified, while the corresponding equation for the semiclassical Green function has been calculated in Refs.[\onlinecite{Zyuzin,Bobkova,Linder}]. The free energy is given by Eq. (\ref{F}). However, in contrast to the Rashba superconductor the second term in Eq.(\ref{F}) is absent.

\section{Vortex coat on a Zeeman island boundary}

The phase $\theta$ of the order parameter is induced by two regular chains of vortices on two opposite edges of a strip at $y=-w/2$ and $y=w/2$, respectively, where $w$ is the strip width. It is assumed that counterclockwise and clockwise vortices with coordinates $\mathbf{r}_i^{+}=(di+s,-w/2,0)$ and $\mathbf{r}_i^{-}=(di,w/2,0)$, are placed on the opposite boundaries, where $d$ is the chain period, $i=1,2,...$, and $s$ denotes the relative shift of the chains in the $x$-direction. In this case $\bm{\nabla}\theta$ in Eq.(\ref{F})  may be written as
\begin{equation}\label{nablatheta}
\bm{\nabla}\theta=\sum_i \left(-\frac{\mathbf{\hat{z}}\times(\mathbf{r}-\mathbf{r}_i^+)}{(\mathbf{r}-\mathbf{r}_i^+)^2}+\frac{\mathbf{\hat{z}}\times(\mathbf{r}-\mathbf{r}_i^-)}{(\mathbf{r}-\mathbf{r}_i^-)^2}\right)\,,
\end{equation}
where $\mathbf{\hat{z}}$ is a unit vector in the $z$-direction. $\bm{\nabla}\theta$ can be averaged over $x$. Let us denote the so averaged $\bm{\nabla}\theta$ as $\overline{\bm{\nabla}\theta}$. From Eq.(\ref{nablatheta}) it is easy to calculate it as $\overline{\bm{\nabla}\theta}=-(2\pi/d)\mathbf{\hat{x}}\theta(x+w/2)\theta(w/2-x)$, where $\mathbf{\hat{x}}$ a unit vector in the $x$-direction. Note, that this average is finite only inside the strip. Therefore, by choosing the appropriate lattice period, one may compensate the Zeeman term in  Eq.(\ref{F}). However, it is necessary to take into account the deviation $\bm{\delta\theta}$ of $\bm{\nabla}\theta$ from its average value. By taking the sum over $i$ in Eq.(\ref{nablatheta}) this deviation can be written in the form
\begin{eqnarray}\label{nablatheta2}
&&\bm{\delta\theta} =\frac{\pi }{d}\sum_{m\neq 0} e^{ig_m x}\left(e^{-|g_m||y-w/2|}\mathbf{V}^+_m-\right.\nonumber\\
&&\left.e^{-|g_m||y+w/2|}e^{-ig_m s}\mathbf{V}^-_m\right)\,,
\end{eqnarray}
where $g_m=2\pi m/d$, $V^{\pm}_x=$sign$(y\mp w/2)$ and $V^{\pm}_y=i$sign$(g_m)$. It is seen that $\bm{\delta\theta}$ decreases fast when the distance from the edges increases. At $2\pi w\gg d $ it contributes in the second term of Eq.(\ref{Fkin}) only within a thin shell $\sim d/2\pi$ near the edges. This term plays a role of the the surface energy. Let us denote it as $\mathcal{F}_{\text{s}}$. By substituting  Eq.(\ref{nablatheta2}) in the second term of  Eq.(\ref{Fkin}) we obtain
\begin{eqnarray}\label{Fs}
&&\mathcal{F}_{\text{s}}=\frac{\pi}{4} N_FD\Delta_0 Lg\left[ 2\ln\frac{2}{g\xi}+\right.\nonumber\\
&&\left.\ln\left(1-e^{-wg+isg}\right)+ \ln\left(1-e^{-wg-isg}\right)\right]\,,
\end{eqnarray}
where $g=2\pi/d$. Since $g\xi \ll 1$, the superconductor coherence length $\xi$ plays the role of the logarithmic cutoff near vortex cores. If the width of the strip is much larger than the distance between vortices the first term in Eq.(\ref{Fs}) dominates and the surface energy does not depend on the strip width. By minimizing the free energy Eq.(\ref{Fkin})  one obtains $d$, as a function of $w$ and $F$. The result is shown in Fig.2.


\begin{thebibliography}{99}
\bibitem{Chandrasekhar}
B. S. Chandrasekhar, Appl. Phys. Lett. \textbf{1}, 7 (1962)
\bibitem{Clogston}
A. M. Clogston, Phys. Rev. Lett. \textbf{9}, 266 (1962)
\bibitem{Fulde}
P. Fulde and R. A. Ferrel, Phys. Rev.  \textbf{135}, A550 (1964)
\bibitem{Larkin}
A. I. Larkin and Yu. N. Ovchinnikov, Zh. Eksp. Teor. Fiz.
\textbf{47}, 1136 (1964) [Sov. Phys. JETP  \textbf{20}, 762 (1965)]
\bibitem{Edelstein}
V. M. Edelstein, Sov. Phys. JETP \textbf{68}, 1244 (1989)
\bibitem{Yip}
S. K. Yip, Phys. Rev. B \textbf{65}, 144508 (2002)
\bibitem{Malsh island}
A.G. Mal'shukov, Phys. Rev. B \textbf{93}, 054511 (2016).
\bibitem{Pershoguba}
S. S. Pershoguba, K. Bj\"{o}rnson, A. M. Black-Schaffer, and A. V. Balatsky, Phys. Rev. Lett. \textbf{115}, 116602 (2015).
\bibitem{Hals}
K. M. D. Hals, Phys. Rev. B \textbf{95}, 134504 (2017)
\bibitem{Rashba}
Yu. A. Bychkov and E. I. Rashba, J. Phys. C \textbf{17}, 6039 (1984).
\bibitem{Samokhin}
V. P. Mineev and K. V. Samokhin, Zh. Eksp. Teor. Fiz. \textbf{105}, 747 (1994) [Sov. Phys. JETP 78, 401 (1994)]
\bibitem{Barzykin}
 V. Barzykin and L. P. Gor'kov, Phys. Rev. Lett. \textbf{89}, 227002 (2002)
\bibitem{Agterberg}
 D. F. Agterberg, Physica C \textbf{387}, 13 (2003)
\bibitem{Kaur}
R. P. Kaur, D. F. Agterberg, and M. Sigrist, Phys. Rev. Lett. \textbf{94}, 137002 (2005)
\bibitem{Agterberg2}
D.F. Agterberg and R.P. Kaur, Phys. Rev. B \textbf{75}, 064511 (2007)
\bibitem{Dimitrova}
O. Dimitrova and M.V. Feigel'man, Phys. Rev. B \textbf{76}, 014522 (2007)
\bibitem{Krive}
 I. V. Krive, A. M. Kadigrobov, R. I. Shekhter and M. Jonson, Phys. Rev. B \textbf{71}, 214516 (2005)
\bibitem{Reinoso}
A. Reynoso, G.Usaj, C.A. Balseiro, D. Feinberg, M.Avignon, Phys. Rev. Lett. \textbf{101}, 107001 (2008)
\bibitem{Zazunov}
 A. Zazunov, R. Egger, T. Martin, and T. Jonckheere, Phys.Rev. Lett. \textbf{103}, 147004 (2009)
\bibitem{ISHE}
A. G. Mal'shukov, S. Sadjina, and A. Brataas, Phys. Rev. B \textbf{81}, 060502 (2010)
\bibitem{Liu}
 J.-F. Liu and K. Chan, Phys. Rev. B \textbf{82}, 125305 (2010)
\bibitem{Yokoyama}
 T. Yokoyama, M. Eto, Y. V. Nazarov, Phys. Rev. B \textbf{89}, 195407 (2014)
\bibitem{Konschelle}
 F. Konschelle,  I. V. Tokatly and F. S. Bergeret, Phys. Rev. B \textbf{92},125443 (2015)
\bibitem{Assouline}
A. Assouline, C. Feuillet-Palma, N. Bergeal, T. Zhang,
A. Mottaghizadeh1, A. Zimmers, E. Lhuillier,
M. Marangolo, M. Eddrief, P. Atkinson, M. Aprili, H. Aubin, Nature Communications
10, 126 (2019)
\bibitem{Malsh Maj}
A. G. Mal'shukov, Phys. Rev. B \textbf{101}, 134514 (2020)
\bibitem{Larkin semiclassical}
A. I. Larkin, and Y. N. Ovchinnikov, Zh. Eksp. Teor. Fiz. \textbf{55}, 2262 (1968) [Sov. Phys. JETP \textbf{28}, 1200 (1965)].
\bibitem{Eilenberger}
G. Eilenberger, Z.Phys. \textbf{214}, 195 (1968)
\bibitem{Rammer}
J. Rammer, H. Smith, Rev. Mod. Phys. \textbf{58}, 323 (1985)
\bibitem{Serene}
J. M. Serene, D. Rainer, Phys. Rep. \textbf{101}, 221 (1983)
\bibitem{Sedlmayr}
N. Sedlmayr, E. W. Goodwin, M. Gottschalk, I. M. Dayton, C. Zhang, E. Huemiller, R. Loloee, T. C. Chasapis, M. Salehi, N. Koirala, M. G. Kanatzidis, S. Oh, D. J. Van Harlingen, A. Levchenko, and S. H. Tessmer, arXiv:180512330
\bibitem{Trang}
C. X. Trang, N. Shimamura, K. Nakayama, S. Souma, K. Sugawara, I. Watanabe, K. Yamauchi, T. Oguchi, K. Segawa, T. Takahashi, Yoichi Ando, and T. Sato, Nat. Communication \textbf{11}, 159 (2020)
\bibitem{Bai}
M. Bai, F. Yang, M. Luysberg, J. Feng, A. Bliesener, G. Lippertz,  A. A. Taskin, J. Mayer, and Y. Ando, arXiv:1910.08331 (2019)
\bibitem{Yoshizawa}
S. Yoshizawa, H. Kim, T. Kawakami, Y. Nagai, T. Nakayama, X. Hu, Y. Hasegawa, and T. Uchihashi, Phys. Rev. Lett. \textbf{113}, 247004 (2014)
\bibitem{Matetskiy}
A. V. Matetskiy, S. Ichinokura, L. V. Bondarenko, A. Y. Tupchaya, D. V. Gruznev, A. V. Zotov, A. A. Saranin, R. Hobara, A. Takayama, and S. Hasegawa, Phys. Rev. Lett. \textbf{115}, 147003 (2015)
\bibitem{Haviland}
D. B. Haviland, Y. Liu, and A. M. Goldman, Phys. Rev. Lett.\textbf{ 62}, 2180 (1989).
\bibitem{Zhang}
T. Zhang, P. Cheng, W. J. Li, Y. J. Sun, G. Wang, X. G. Zhu, K. He, L. L.Wang, X. C. Ma, X. Chen, Y. Y.Wang,
Y. Liu, H. Q. Lin, J. F. Jia, and Q. K. Xue, Nat. Phys. \textbf{6}, 104 (2010).
\bibitem{Uchikashi}
T. Uchihashi, P. Mishra, M. Aono, and T. Nakayama, Phys. Rev. Lett. \textbf{107}, 207001 (2011).
\bibitem{Sekihara}
T. Sekihara, R. Masutomi, and T. Okamoto, Phys. Rev. Lett. \textbf{111}, 057005 (2013)
\bibitem{Pavlov}
D. P. Pavlov, R. R. Zagidullin, V. M. Mukhortov, V. V. Kabanov, T. Adachi, T. Kawamata, Y. Koike, and R. F. Mamin,  Phys. Rev. Lett. \textbf{122}, 237001 (2019)
\bibitem{Caviglia}
 A. D. Caviglia, M. Gabay, S. Gariglio, N. Reyren, C. Cancellieri, and J. M. Triscone, Phys. Rev. Lett. \textbf{104}, 126803 (2010)
\bibitem{Houzet}
Manuel Houzet and Julia S. Meyer, Phys. Rev. B \textbf{92}, 014509 (2015).
\bibitem{Zyuzin}
A. Zyuzin, M. Alidoust, and D. Loss, Phys. Rev. B \textbf{93}, 214502 (2016).
\bibitem{Bobkova}
I. V. Bobkova, A. M. Bobkov, A. A. Zyuzin, and M.Alidoust, Phys. Rev. B \textbf{94}, 134506 (2016)
\bibitem{Linder}
Henning G. Hugdal, Jacob Linder, and Sol H. Jacobsen, Phys. Rev. B 95, 235403 (2017)
\bibitem{Pearl}
J. Pearl, Appl. Phys. Lett, \textbf{5}, 65 (1964).
\bibitem{AGD}
A. A. Abrikosov, L. P. Gor'kov, and I. E. Dzyaloshinskii, Methods
of Quantum Field Theory in Statistical Physics (Dover, New York, 1975)
\bibitem{Fu}
L. Fu and C. L. Kane, Phys. Rev. Lett. \textbf{100}, 096407 (2008)
\bibitem{Jackiw}
R. Jackiw and P. Rossi, Nucl. Phys. B \textbf{190}, 681 (1981).
\bibitem{Read}
N. Read and D. Green, Phys. Rev. B \textbf{61}, 10267 (2000)
\bibitem{Ivanov}
D. A. Ivanov, Phys. Rev. Lett. \textbf{86}, 268 (2001).
\bibitem{Usadel}
K. D. Usadel, Phys. Rev. Lett.\textbf{ 25}, 507 (1970).
\bibitem{Franz}
M. Franz and Te\~{s}anovi\'{c}, Phys. Rev. Lett, \textbf{84}, 554 (2000)
\end{thebibliography}
\end{document}